\documentstyle {article}
\begin{document}
\title{Intersecting Braids and Intersecting Knot Theory}
\author{{\small Daniel Armand-Ugon, Rodolfo Gambini and Pablo
Mora}\\
{\small Instituto de F\'{\i}sica, Facultad de Ciencias }\\
{\small Trist\'an Narvaja 1674, Montevideo,Uruguay}}
\date{\today}
\maketitle
\begin{abstract}
An extension of the Artin Braid Group with new operators that
generate double and triple intersections is considered.
The extended Alexander theorem, relating intersecting closed
braids and
intersecting knots is proved for double
and triple intersections, and a
counter example is given for the case of quadruple
intersections.
Intersecting knot invariants are constructed via Markov traces
defined on intersecting
braid
algebra representations, and the extended Turaev representation
is discussed
as
an
example.\\
Possible applications of the formalism to quantum gravity are
discussed.
\end{abstract}

{\bf 1. Introduction}\\

During the last years a great deal of interest has been paid
to the
construction and study of intersecting
link invariants. This motivation is due to the important role that
intersecting
links seem to play in several fields.\\
Beginning with the construction by V. Vassiliev \cite{vas} of
 the so called 'Vassiliev invariants'
of knots via singularity theory. Vassiliev considered not the knot
itself, but
the map from $S^1$ to $S^3$ that generates the knot.
Within this approach,
different
knots correspond to the  disconnected
components of the space of isotopy classes of these maps (those
which are
 embeddings).
The 'walls' among the components are then knots with intersections
and other
types of
of singularities. The construction
by Vassiliev of a recursive (skein-type) procedure to compute the
invariants
involves
as an intermediate step a going trough these walls, and thus the
Vassiliev
invariants
take
values on intersecting links, which appear then as naturally as
non intersecting links.\\
A a second example, the Witten's formula \cite{wit}  relating
Chern-Simons
field
theory and knot invariants was originally derived from conformal
field theory.
In alternative derivations  via the so called
variational approach, singular invariants are involved, since
\cite{guad,gampul,kau3,agm} the derivation of  the
skein relation is done by
the breaking of a double intersection. Furthermore some of these
works
\cite{gampul,agm} support
the viewpoint that  Witten's formula also  holds for double and
triple
intersections,
and that there are no natural distinction between the intersecting
and the
non intersecting case.\\
As a final example, in the loop representation of Quantum Gravity
\cite{rov}
the physical
 states are
functionals of the knot classes, ie. knot invariants. It has also
been proved
that for nondegenerate solutions (those for which the spatial
volume
does not vanish)
 the wave function take non-zero values
on  knots with at least triple intersections
\cite{husain,gampul}.\\
Our motivation in this paper comes from our work in Quantum Gravity,
where physical states must be invariant under diffeomorfhisms.
This leads us to consider a slight shift of the standard conceptual
frame of
knot theory. We will consider diffeomorphisms connected
to the identity on the
manifold ($R^3$ or $S^3$ in this work) instead of homeomorfisms, as
usual.
This makes no difference for nonintersecting knot theory, and it is
implicitly
assumed in previous works about double intersections.
In the tangent space at any point
the diffeomorphims induces an invertible lineal transformation, that
preserves
the linear independence (dependence) between tangent vectors.
Furthermore a diffeomorphism connected to the identity preserves
the sign of
the volume spanned by three arbitrary vectors in the tangent space.\\
This paper is organized as follows: in section 2 we consider an
extension of
 the braid algebra for double
and triple intersections. In section 3 matrix representations
of the extended
algebra are considered.
Finally in section 4
we show how  intersecting link invariants can be obtained, using
these
 representations
or using arbitrary non intersecting invariants, no matter  how
they were derived.\\
 We would like to point out that the rigor level is that of a
 theoretical
 physicist.\\

{\bf 2. New Reidemeister moves, extended braid algebra and the
Alexander
theorem}\\

In this section we will extend the usual Reidemeister moves planar
knots (links) diagrams to include double and triple intersections.\\

Later we will
prove that these generalized knots (links) may be drawn as closed
braids
(generalized Alexander theorem). Finally we enlarge the braid
algebra by
including new elements that generate double and triple
intersections.\\

The Reidemeister moves for double intersections, that we will call D,
are sketched in figure 1.
This kind of intersections and their associated moves were considered
first
in \cite{kau1} and later in many works
\cite{gampul,kau2,bae,bir1,Smo}.
For the case of triple intersections\cite{agm} we must note that
the sign of
 the volume element
spanned by the tangents at the triple point is an invariant under
a diffeomorphism connected with the identity. Hence there are three
classes of triple points, that we will call $T^+$, $T^-$ and $T^0$.
These
intersections and their Reidemeister moves are sketched in
figures 2 and 3.\\
Note that for triple intersections we may assume (after an isotopy)
that
our incoming and outcoming
lines in the diagram are arranged so there are no incoming
(outcoming)
lines in between two outcoming (incoming) lines. The only
exception is the
flat intersection drawn in figure 4, that we will ignore
in what follows.
Due to this fact one can show that except for the above mention
case, the
Alexander theorem \cite{Al} still holds ( for double intersections
this
was shown by J. Birman
\cite{bir1}). \\
{\bf Theorem 1}. {\em Given an arbitrary representative K of a
intersecting knot
or link with double and triple points
and an axis A in} $R^3-K$ {\em ,K may be deformed to a closed
intersecting
braid with axis A}.\\
{\em Proof.} If we think of  A as the $z$ axis, by making an
isotopy of K
in $R^3-K$ we may put it in the $(r,\theta )$ plane (except
in a small
neighborhood of a crossing or a $T^{+,-}$ intersection). By
a another
isotopy we may arrange that the polar angle function restricted
to a small neighborhood of each intersection be monotonically
increasing. We now proceed as in the usual proof \cite{Al,bir1} by
considering the arcs of K that have at most one crossing and have
not intersecting
points, and classificating these arcs as good (bad) if the polar
angle function
is an increasing (decreasing) function of the arc. Replacing the
bad arcs
(if there are) for good arcs by deforming K we finally get a closed
intersecting braid $\Box$ \\
Note that the crucial step in the proof is the possibility of
arranging the
lines at an intersecting point so that between two incoming
(outcoming) lines
there are no outcoming (incoming) lines. This is not possible
for one
of the two classes of flat triple intersection and several classes
of
quadruple intersections. In fig.4  examples of this are shown.
These  intersections , if they appear in a knot, prevent  it from
being expressed
as a closed braid.

We can conclude that a knot with n-uple intersections, $n>3$,
cannot be
expressed generally as a closed braid, despite of the fact that
any set of
three tangents at the
intersecting point are lineary independent.\\

An algebraic expression of the topological properties of  braids
is given by
the Artin braid group $B_N$ that is generated by $N$ elements $g_i$,
$i=1,...,N-1$
satisfying the algebra
\begin{eqnarray}
&&g_ig_i^{-1}=g_i^{-1}g_i=I,\;\;\;\;g_ig_{i+1}g_i=
g_{i+1}g_i g_{i+1} \nonumber
\\
&&g_ig_j=g_jg_i\;\;\;if\;\mid i-j\mid >1
\end{eqnarray}
The braid algebra $B_N$ can be extended to include double
\cite{kau1,Smo,bae,bir1} and triple
intersections \cite{agm} by adding the generators $a_j$, $b_i^+$,
$b_i^-$ and
$b_i^0$, $j=1,...,N-1$, $i=2,...,N-1$ (see figure 5).
In what follows we call $SB_N$ the extended braid algebra.
These generators must
satisfy algebraic relations, that follow from the generalized
Reidemeister
moves of fig.1 and fig.3. They are
\begin{eqnarray}
&&a_ig_i=g_ia_i,\;\;\; g_i^{-1}a_{i+1}g_i=
g_{i+1}a_ig_{i+1}^{-1}\nonumber \\
&&g_ia_j=a_jg_i\;\;\; and\;\;\;a_ia_j=a_ja_i\;\; if\; \mid i-j\mid >1
\end{eqnarray}
and
\begin{eqnarray}
&&b_i^-=g_{i-1}b_i^+g_i^{-1},\;\; b_i^-=g_i^{-1}b_i^+g_{i-1},\;\;
b_{i+1}^{+,-,0}=g_{i-1}g_ig_{i+1}b_i^{+,-,0}
g_{i+1}^{-1}g_i^{-1}g_{i-1}^{-1}\nonumber \\
&&b_ig_j=g_jb_i\;\;and\;\;b_ia_j=a_jb_i\;\; if\;\; j>i+1\;\; or
\;\; j<i-2
\nonumber \\
&&b_ib_j=b_jb_i\;\; if\;\;\mid i-j\mid
>2,\;\;b_i^0=g_ig_{i-1}g_ib_i^0g_i^{-1}g_{i-1}^{-1}g_i^{-1}.
\end{eqnarray}
We would like to point  out that the extended braid algebra is
not a group,
because
the new generators have no inverse.\\
The fact that knots with n-uple intersections cannot, in general,
be expressed as  braids makes it less interesting further extensions
of
the braid algebra\footnote{Perhaps the way to do this is to enlarge
the
Temperley-Lieb algebra}. There exist other classes of double and
triple
intersections for which the incoming and outcoming tangents to any
of the strands
are not the same (there are kinks at the intersection).
These cases will not be discussed here but are relevant in quantum
gravity \cite{mor}.
\\

{\bf 3. Representations of the extended braid algebra}\\

We will consider representations of the braid algebra of the $g_i$'s
by complex matrices $G_i$ satisfying eq.(1). To construct a
representation of the extended braid algebra we must  add
matrices $A_i$ \cite{gampul}, $B_i^{\pm}$ and $B_i^0$ \cite{agm}
satisfying
eq.(2-3).
The general procedure would be, given a representation of the
standard braid algebra, to impose eq.(2-3) in its matrix form.
We can see, however, that matrices of the form
\begin{eqnarray}
&&A_i=\alpha _1G_i+\alpha _2G_i^{-1}+\alpha _3I_i \nonumber \\
&&B_i^+=\beta _1G_iG_{i-1}G_i+\beta _2
(G_iG_{i-1}+G_{i-1}G_i+G_i^2)+\beta _3
G_i+\beta _4G_{i-1}^{-1} \nonumber \\
&&B_i^-=\beta _1G_iG_{i-1}G_i+\beta _2(G_iG_{i-1}+G_{i-1}G_i+
G_{i-1}^2)+\beta
_3
G_{i-1}+\beta _4G_{i}^{-1} \nonumber \\
&&B_i^0=\beta _5[B_i^++B_i^-]+\beta _6I_{i-1}I_i
\end{eqnarray}
are solution of eq.(2-3). Thus, given a representation of the
braid group
it is straightforward to construct a representation of the
extended braid
algebra.\\
A wide class of $B_N$ representations are the so called
R-matrix
representations, where a finite dimensional vector space $V_i$ is
associated
to the i-th strand and the total representation space is
$V(N)=V_1\otimes .\;
.\; .\; .\otimes V_N$.
The $G_i$'s are given by
\begin{equation}
G_i=I\otimes\;.\; .\; .\otimes R\otimes\; .\; .\; .\otimes I
\end{equation}
where R acts on $V_i\otimes V_{i+1}$. To have a representation
of $B_N$ we
 require that $R$ be invertible and
\begin{equation}
R_{12}R_{23}R_{12}=R_{23}R_{12}R_{12}
\end{equation}
where
\begin{equation}
R_{12}=R\otimes I \; ,\; R_{23}=I\otimes R
\end{equation}
The eq.(6) is called the quantum Yang-Baxter equation.\\
To enlarge the $B_N$ representation to a representation of the
extended braid
algebra  $SB_N$
we must  add the new generators given by
\begin{eqnarray}
&&A_i=I\otimes\; .\; .\; .\otimes A\otimes\; .\; .\; .\otimes
I \nonumber \\
&&B_i^{\pm ,0}=I\otimes\; .\; .\; .\otimes B^{\pm ,0}\otimes\; .\; .
\;
.\otimes I
\end{eqnarray}
where $A$ acts on $V_i\otimes V_{i+1}$ and $B^{\pm ,0}$ acts on
$V_{i-1}\otimes V_i\otimes V_{i+1}$
and to impose the relations given by Eq (2-3).
\begin{equation}
AR=RA\; ,\; A_{23}R_{12}R_{23}=R_{12}R_{23}A_{12}
\end{equation}
with
\begin{equation}
A_{12}=A\otimes I ,\;\; A_{23}=I\otimes A
\end{equation}
and
\begin{eqnarray}
B^-=R_{12}B^+R_{23}^{-1},\;\; B^-=R_{23}^{-1}B^+R_{12},\;\;
B^0=R_{23}R_{12}R_{23}B^0R_{23}^{-1}R_{12}^{-1}R_{23}^{-1} \nonumber
\end{eqnarray}
\begin{eqnarray}
I\otimes B^{\pm ,0}&=&(R_{12}\otimes I)(I\otimes R_{12})(I\otimes
R_{23})
(B^{\pm ,0}\otimes I)\times \nonumber\\
&&\times(R_{12}\otimes I)^{-1}(I\otimes R_{12})^{-1}(I\otimes
R_{23})^{-1}
\end{eqnarray}
We explicitly worked out the computations involved in eq.(9-11)
for the Turaev
representation of $SB_N$. In the Turaev  representation the space
$V_i$ is
the two dimensional complex vector space $C^2$ with basis vectors
$\{ u_i,v_i\}$, and the R-matrix is
\begin{eqnarray}
R=q^{\frac{1}{4}}\left(\begin{array}{cccc}
1 & 0 & 0 & 0 \\
0 & 1-q^{-1} & q^{-\frac{1}{2}}& 0 \\
0 & q^{-\frac{1}{2}} & 0 & 0 \\
0 & 0 & 0 & 1
\end{array}\right)
\end{eqnarray}in the basis $\{ u_i\otimes u_{i+1},u_i\otimes v_{i+1}
,v_i\otimes u_{i+1},v_i\otimes v_{i+1}\}$ of $V_i\otimes V_{i+1}$.
The complex parameter $q$ is arbitrary and the generators in this
representation satisfy the relation
\begin{equation}
q^{\frac{1}{4}}G_i-q^{-\frac{1}{4}}G_i^{-1}=(
q^{\frac{1}{2}}-q^{-\frac{1}{2}})
I
\end{equation}
that leads to the skein relations. \\
Our direct calculation for A showed that the general form of this
matrix is
given by eq.(4), that is
\begin{equation}
A=\alpha _1 R+\alpha _2 R^{-1}
\end{equation}
where the $\alpha$'s are arbitrary complex parameters.
In eq.(14) $\alpha_3 I$ does not appear  because from eq.(13) we
know that
$I$ is a linear combination of $G_i$ and $G_i^{-1}$. The explicit
computation
 of
$B^+$, $B^-$ and $B^0$ is again in accordance with eq.(4), the
$\beta$'s not
being all
independent. The general result in this representation is
\begin{eqnarray}
&&B^+=\beta _1 M_1 +\beta _3M_3+\beta _4M_4\nonumber\\
&&B^-=\beta _1 M_1 +\beta _3M^{-1}_4+\beta _4M^{-1}_3\nonumber\\
&&B^0=\beta _5[B^++B^-]+\beta _6I
\end{eqnarray}
with
\begin{eqnarray}
&&M_1=(I\otimes R)(R\otimes I)(I\otimes R)\nonumber\\
&&M_3=(I\otimes R)\nonumber\\
&&M_4=(R^{-1}\otimes I)
\end{eqnarray}
or explicitly

\begin{eqnarray}
M_1=\pmatrix{ {q^{{3\over 4}}} & 0 & 0 & 0 & 0 & 0 & 0 & 0 \cr 0 &
  {{-{q^{{5\over 4}}} + {q^{{9\over 4}}}}\over {{q^{{3\over 2}}}}} &
  -{q^{-{3\over 4}}} + {q^{{1\over 4}}} & 0 & {q^{-{1\over 4}}} & 0
  & 0 & 0
   \cr 0 & -{q^{-{3\over 4}}} + {q^{{1\over 4}}} & {q^{-{1\over 4}}}
   & 0 & 0
   & 0 & 0 & 0 \cr 0 & 0 & 0 & {{-{q^{{1\over 4}}} +
    {q^{{5\over 4}}}}\over
    {{\sqrt{q}}}} & 0 & -{q^{-{3\over 4}}} + {q^{{1\over 4}}} &
  {q^{-{1\over 4}}} & 0 \cr 0 & {q^{-{1\over 4}}} & 0 & 0 & 0 & 0
  & 0 & 0 \cr
  0 & 0 & 0 & -{q^{-{3\over 4}}} + {q^{{1\over 4}}} & 0 &
  {q^{-{1\over 4}}} &
  0 & 0 \cr 0 & 0 & 0 & {q^{-{1\over 4}}} & 0 & 0 & 0 & 0 \cr 0 & 0
  & 0 & 0 &
  0 & 0 & 0 & {q^{{3\over 4}}} \cr  }
 \end{eqnarray}

 \begin{eqnarray}
M_3=\pmatrix{ {q^{{1\over 4}}} & 0 & 0 & 0 & 0 & 0 & 0 & 0 \cr 0 &
  {{-{q^{{3\over 4}}} + {q^{{7\over 4}}}}\over {{q^{{3\over 2}}}}} &
  {q^{-{1\over 4}}} & 0 & 0 & 0 & 0 & 0 \cr 0 & {q^{-{1\over 4}}} &
  0 & 0 & 0
   & 0 & 0 & 0 \cr 0 & 0 & 0 & {q^{{1\over 4}}} & 0 & 0 & 0 & 0 \cr 0
   & 0 & 0
   & 0 & {q^{{1\over 4}}} & 0 & 0 & 0 \cr 0 & 0 & 0 & 0 & 0 &
  {{-{q^{{3\over 4}}} + {q^{{7\over 4}}}}\over {{q^{{3\over 2}}}}} &
  {q^{-{1\over 4}}} & 0 \cr 0 & 0 & 0 & 0 & 0 & {q^{-{1\over 4}}} &
  0 & 0 \cr
  0 & 0 & 0 & 0 & 0 & 0 & 0 & {q^{{1\over 4}}} \cr  }
 \end{eqnarray}

 \begin{eqnarray}
M_4=\pmatrix{ {q^{-{1\over 4}}} & 0 & 0 & 0 & 0 & 0 & 0 & 0 \cr 0 &
  {q^{-{1\over 4}}} & 0 & 0 & 0 & 0 & 0 & 0 \cr 0 & 0 & 0 & 0 &
  {q^{{1\over 4}}} & 0 & 0 & 0 \cr 0 & 0 & 0 & 0 & 0 &
  {q^{{1\over 4}}} & 0 &
  0 \cr 0 & 0 & {q^{{1\over 4}}} & 0 & {{1 - q}
  \over {{q^{{1\over 4}}}}} & 0
   & 0 & 0 \cr 0 & 0 & 0 & {q^{{1\over 4}}} & 0 &
  {{{q^{{5\over 4}}} - {q^{{9\over 4}}}}\over {{q^{{3\over 2}}}}} &
  0 & 0 \cr
  0 & 0 & 0 & 0 & 0 & 0 & {q^{-{1\over 4}}} & 0 \cr 0 & 0 & 0 & 0 &
  0 & 0 & 0
   & {q^{-{1\over 4}}} \cr  }
 \end{eqnarray}

Note that we omit the $\beta _2$ term because it may be expressed as
linear combination of the others. It  could happen that the most
general
B's are of the form given by eq.(4)  plus other terms irreducible
to this
form.
If this were the case,
 we would not have simple skein relations for our intersecting knot
invariants, as will be shown later.\\

{\bf 4. From intersecting braids to intersecting knot invariants}\\

In this section we will show how to construct intersecting link
invariants from intersecting braids. \\
As we already saw, any oriented intersecting link of the classes
that
we considered can be expressed as a closed intersecting braid.
Conversely
given any braid diagram we can associate a link diagram by joining
the
endpoints
of the strands in an order preserving way and without adding new
crossings.
We call  the resulting link the closure of this braid element
$b\epsilon SB_N$
and we denote it as $\hat{b}$. It is straightforward to see that
the extended
braid algebra relations account for the type 2 and 3  Reidemeister
as well as
for
the new moves. But we also have  that braids related by the
following moves
(Markov moves) correspond to ambient isotopic links \cite{mar}:\\
M1. $b_1,b_2\in B_N,\rightarrow\widehat{b_1b_2}
\equiv\widehat{b_1b_1}$\\
M2. $b_1\in B_N,\;\; b_2=b_1g^{\pm 1}_{N+1}\in
B_{N+1}\rightarrow\widehat{b_1}\equiv\widehat{b_2}$\\
If we are working with regular isotopy only the first type of Markov
moves
holds.\\
To construct intersecting links invariants from an R-matrix
representation of
the
extended braid algebra we will follow the standard procedure
of Markov traces.
It has been proved \cite{Turaev} that if there is a matrix
$\mu$ acting in
$V_i$ such that
\begin{eqnarray}
&&R\mu _1\otimes\mu _2=\mu _1\otimes\mu _2R\nonumber\\
&&Tr\mid _{V_2}(R^{\pm 1}\mu _2)=\alpha ^{\pm 1}.I\mid_{V_i}
\end{eqnarray}
then, if $L$ is the link obtained from the closure $\widehat{b}$
of the braid
$b\in B_N$ represented by the matrix $B$, we have that
\begin{equation}
\chi ^R(L)=Tr\mid _{V(N)}(B\mu ^{\otimes N})
\end{equation}
is a regular isotopy invariant and that
\begin{equation}
\chi ^A(L)=\alpha ^{-w(L)}Tr\mid _{V(N)}(B\mu ^{\otimes N})
\end{equation}
 where $w(L)$ is the writhe of the diagram, is an ambient isotopy
 invariant.
Since the new Reidemeister moves are accounted for in the extended
algebra, it
is
straightforward to see that invariants for intersecting links can
be defined
in the same way. This allows us to extend known link invariants
(for regular
or
ambient isotopy) to the intersecting case.
In general we will have, in addition of the usual defining
relations for the non intersecting case (usually skein relations
and the value
of
the unknotted), the generalized skein relations
\begin{eqnarray}
\chi (D)&=&\alpha _1\chi (L_+)+\alpha _2\chi (L_-)+
\alpha _3\chi (L_0)
\nonumber \\
\chi (T^+)&=&\beta _1\chi (L_{2+}L_{1+}L_{2+})+\beta _2[\chi (L_{1+}
L_{2+})+
\chi (L_{2+}L_{1+})+\chi (L_{2+}L_{2+})]+\nonumber\\
&&+\beta _3\chi (L_{2+})+\beta _4\chi (L_{1-}) \nonumber \\
\chi (T^-)&=&\beta _1\chi (L_{2+}L_{1+}L_{2+})+\beta _2[\chi (L_{1+}
L_{2+})+
\chi (L_{2+}L_{1+})+\chi (L_{1+}L_{1+})]+\nonumber\\
&&+\beta _3\chi (L_{1+})+\beta _4\chi(L_{2-}) \nonumber \\
\chi (T^0)&=&\beta_5[\chi (T^+)+\chi (T^-)]+\beta _6\chi (L_{10}
L_{20})
\end{eqnarray}
where the meaning of the notation is explained in fig.6.\\
Eq.(23) follows from eq.(4), eq.(21-22) and the properties of the
trace. For
example,
for the generalized Turaev representation and with the matrix $\mu$
\begin{eqnarray}
\mu =
\left(\begin{array}{cc}
q^{-\frac{1}{2}}&0\\
0&q^{\frac{1}{2}}
\end{array}\right)
\end{eqnarray}

we obtain a generalized Jones polynomial\cite{jon} in the variables
$q,\;\alpha _i,\;\beta _j$.\\

The approach via extended braid algebra and Markov traces makes
crystal clear
our construction of intersecting links invariant, and allows to
detect
redundancy
in the chosen parameters. However a direct (and more general)
construction
is possible.\\
{\bf Theorem 2.} {\em Given any ambient (regular) isotopic invariant
 for
non-intersecting links $\chi (L)$ its generalization to the
intersecting
case $\chi ^{E}(L)$ given by eq.(23) is an intersecting link
invariant}\\
{\em Proof.} From the defining relations of eq.(23), making a
sequence of
transformations on planar diagrams entirely analogous to the
algebraic
calculations needed to check that the matrices of eq.(4) satisfy
the algebra
(2-3),
and using that $\chi (L)$ is a non-intersecting link invariant i.e.
invariant under type 2 and 3 Reidemeister moves, the invariance
 of $\chi ^E(L)$ under the new moves can be verified, that concludes
the proof $\Box$.\\
Note that the defining relations eq.(23) imply that the extended
invariants
inherit many properties from the non-intersecting original ones.
For example
the
polynomials defined above from R-matrix representations and Markov
traces
don't detect the orientation of the links, despite of its appearance,
because
their original ones do not do it either.\\

{\bf 5. Concluding remarks and open problems}\\

In the study of Vassiliev invariants only double intersections have
been
 considered (these are unavoidable). The skein-type relation used
is of the form of eq.(23) with $\alpha _1=-\alpha _2$ and
$\alpha _3=0$
(see for example \cite{bir1,bae}).
It is interesting to ask what happens for triple (or n-uple)
intersections and to study if we also have relations of the
form given by
eq.(23).\\
For Chern-Simons theory the work in the intersecting case is quite
heuristic;
it would be desirable to have a rigorous proof of the validity of
Witten's
formula in the intersecting case.\\
In loop quantum gravity,
some physical wave functions are given in integral
form, an  example being the Gauss linking number
$$\chi
(L_1,L_2)=\frac{1}{4\pi}\oint_{L_1}dx^{\mu}\oint_{L_2}dy^{\nu}
\epsilon_{\mu\nu
\rho}
\frac{(x-y)^{\rho}}{\mid x-y\mid ^3}$$
To have a nondegenerate solution we need it to take  non-zero
values
on at  least triple intersections, but the integral is ill defined
in this case. The usual procedure is to consider a framing like
$x^{\mu}\rightarrow x^{\mu}+\delta^{\mu}(x)$ and an analogous one
to $y^{\nu}$
that splits the intersections. Theorem 2 tell us how the framing
must
be taken. Not all the  framings work.  We will always have an
invariant of
the framed links, but in general this will not be the same as
having an
invariant of the original intersecting loop.\\
On the other hand it has been proved that link invariants with
non-zero
values only on non intersecting links are degenerate physical
states of
loop quantum gravity \cite{rov}, provided they do not detect
the orientation
of the link. In the intersecting case the so called hamiltonian
constraint
imposes new conditions. For example, if the cosmological constant
$\Lambda$
does not vanish, we have that the Kauffman Bracket
$S(L;q,\alpha _i,\beta_j)$
(see for example \cite{kau3}), a polynomial invariant
closely related to the Jones polynomial, is a physical
state $\psi _{\Lambda}[L]$
\cite{gampul,agm}, where the
parameters $q,\;\alpha _i$ and $\beta _j$ are fixed
functions of $\Lambda$.
That
is: $\psi _{\Lambda}[L]=S(L;q(\Lambda),\alpha _i(\Lambda),
\beta _j(\Lambda))$.
One can think that the hamiltonian does select one 'point'
of the 'parameter
space'
spanned by $q,\;\alpha _i$ and $\beta _j$.\\
In principle, each non intersecting invariant, that do not
detect the orientation,
may be extended to the intersecting case in such a way that
the Mandelstam constraints
of Quantum Gravity are satisfyed.
This raises the question \cite{mor} of what values of the $\alpha 's$
and $\beta 's$
are required in order to have a non degenerate physical state of
quantum
gravity for a given value of $\Lambda$.\\

\newpage

\unitlength=0.50mm
\special{em:linewidth 0.4pt}
\linethickness{0.4pt}
\begin{picture}(170.00,60.00)
\
\put(35.00,40.00){\vector(0,1){20.00}}
\
\put(5.00,40.00){\vector(0,1){20.00}}
\put(20.00,30.22){\makebox(0,0)[cc]{$\bullet$}}
\
\put(50.00,44.89){\vector(0,1){15.11}}
\
\put(80.00,44.89){\vector(0,1){15.11}}
\put(105.00,0.00){\vector(0,1){60.00}}
\
\put(125.00,40.00){\vector(0,1){20.00}}
\
\put(115.00,40.00){\vector(0,1){20.00}}
\
\put(160.00,40.00){\vector(0,1){20.00}}
\
\put(150.00,40.00){\vector(0,1){20.00}}
\
\put(140.00,50.22){\vector(0,1){9.78}}
\
\put(155.00,30.22){\makebox(0,0)[cc]{$\bullet$}}
\put(120.00,30.22){\makebox(0,0)[cc]{$\bullet$}}
\put(65.00,28.00){\makebox(0,0)[cc]{$\bullet$}}
\put(135.00,30.22){\vector(1,0){5.00}}
\put(42.00,30.22){\vector(1,0){7.00}}
\put(42.00,30.22){\vector(-1,0){7.00}}
\put(135.00,30.22){\vector(-1,0){6.00}}
\put(5.00,0.00){\line(0,1){20.00}}
\put(5.00,20.00){\line(3,2){30.00}}
\put(35.00,0.00){\line(0,1){20.00}}
\put(35.00,20.00){\line(-3,2){30.00}}
\put(50.00,0.00){\line(0,1){10.22}}
\put(50.00,10.22){\line(3,1){30.00}}
\put(80.00,20.00){\line(-2,1){30.00}}
\put(50.00,35.11){\line(3,1){13.00}}
\put(69.00,40.89){\line(3,1){11.00}}
\put(115.00,0.00){\line(0,1){20.00}}
\put(115.00,20.00){\line(1,2){9.67}}
\put(125.00,0.00){\line(0,1){20.00}}
\put(125.00,20.00){\line(-1,2){9.00}}
\put(140.00,0.00){\line(0,1){10.22}}
\put(140.00,10.22){\line(3,1){9.00}}
\put(152.00,14.22){\line(3,1){6.00}}
\put(158.00,16.00){\line(3,1){3.00}}
\put(161.00,16.89){\line(3,1){9.00}}
\put(170.00,20.00){\line(0,1){20.00}}
\put(170.00,40.00){\line(-3,1){9.00}}
\put(161.00,43.11){\line(-3,1){3.00}}
\put(158.00,44.00){\line(-5,2){6.00}}
\put(149.00,47.11){\line(-3,1){9.00}}
\put(150.00,0.00){\line(0,1){20.00}}
\put(150.00,20.00){\line(1,2){10.00}}
\put(160.00,0.00){\line(0,1){20.00}}
\put(160.00,20.00){\line(-1,2){10.00}}
\put(80.00,0.00){\line(0,1){10.22}}
\put(80.00,10.22){\line(-3,1){12.00}}
\put(62.00,16.00){\line(-3,1){12.00}}
\put(50.00,20.00){\line(2,1){30.00}}
\put(80.00,35.11){\line(-3,1){30.00}}
\put(116.00,38.22){\line(-3,5){1.00}}
\put(93.00,30.22){\makebox(0,0)[cc]{and}}
\end{picture}\\
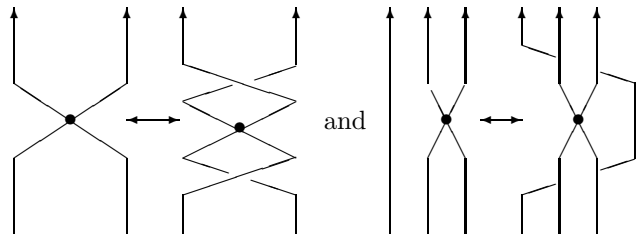
{\bf Figure 1.}Reidemeister moves for a double intersection. \\

\unitlength=0.50mm
\special{em:linewidth 0.4pt}
\linethickness{0.4pt}
\begin{picture}(165.00,50.22)
\put(5.00,4.89){\line(0,1){15.11}}
\put(5.00,20.00){\line(0,1){24.89}}
\put(5.00,44.89){\line(1,0){40.00}}
\put(45.00,44.89){\line(0,-1){40.00}}
\put(45.00,4.89){\line(-1,0){40.00}}
\put(10.00,0.00){\line(0,1){10.22}}
\put(10.00,10.22){\line(1,1){30.00}}
\put(40.00,40.00){\vector(0,1){10.22}}
\put(40.00,0.00){\line(0,1){10.22}}
\put(40.00,10.22){\line(-1,1){30.00}}
\put(10.00,40.00){\vector(0,1){10.22}}
\put(25.00,0.00){\line(0,1){24.89}}
\put(25.00,26.22){\line(0,1){2.67}}
\put(25.00,32.00){\line(0,1){3.11}}
\put(25.00,38.22){\line(0,1){2.67}}
\put(25.00,43.11){\line(0,1){1.78}}
\put(25.00,46.22){\vector(0,1){4.00}}
\put(65.00,4.89){\line(0,1){15.11}}
\put(65.00,20.00){\line(0,1){24.89}}
\put(65.00,44.89){\line(1,0){40.00}}
\put(105.00,44.89){\line(0,-1){40.00}}
\put(105.00,4.89){\line(-1,0){40.00}}
\put(70.00,0.00){\line(0,1){10.22}}
\put(70.00,10.22){\line(1,1){30.00}}
\put(100.00,40.00){\vector(0,1){10.22}}
\put(100.00,0.00){\line(0,1){10.22}}
\put(100.00,10.22){\line(-1,1){30.00}}
\put(70.00,40.00){\vector(0,1){10.22}}
\put(85.00,43.11){\line(0,1){1.78}}
\put(85.00,46.22){\vector(0,1){4.00}}
\put(125.00,4.89){\line(0,1){15.11}}
\put(125.00,20.00){\line(0,1){24.89}}
\put(125.00,44.89){\line(1,0){40.00}}
\put(165.00,44.89){\line(0,-1){40.00}}
\put(165.00,4.89){\line(-1,0){40.00}}
\put(130.00,0.00){\line(0,1){10.22}}
\put(130.00,10.22){\line(1,1){30.00}}
\put(160.00,40.00){\vector(0,1){10.22}}
\put(160.00,0.00){\line(0,1){10.22}}
\put(160.00,10.22){\line(-1,1){30.00}}
\put(130.00,40.00){\vector(0,1){10.22}}
\put(145.00,0.00){\line(0,1){24.89}}
\put(145.00,43.11){\line(0,1){1.78}}
\put(145.00,46.22){\vector(0,1){4.00}}
\put(85.00,24.89){\line(0,1){17.33}}
\put(85.00,0.00){\line(0,1){4.89}}
\put(85.00,8.00){\line(0,1){4.00}}
\put(85.00,15.11){\line(0,1){4.00}}
\put(85.00,22.22){\line(0,1){2.67}}
\put(145.00,24.89){\line(0,1){17.33}}
\put(149.00,24.89){\makebox(0,0)[cc]{0}}
\put(160.00,24.89){\makebox(0,0)[cc]{$T^0$}}
\put(100.00,24.89){\makebox(0,0)[cc]{$T^-$}}
\put(89.00,24.89){\makebox(0,0)[cc]{-}}
\put(40.00,24.89){\makebox(0,0)[cc]{$T^+$}}
\put(29.00,24.89){\makebox(0,0)[cc]{+}}
\put(25.00,24.89){\makebox(0,0)[cc]{$\bullet$}}
\put(85.00,24.89){\makebox(0,0)[cc]{$\bullet$}}
\put(145.00,24.89){\makebox(0,0)[cc]{$\bullet$}}
\end{picture}
\\
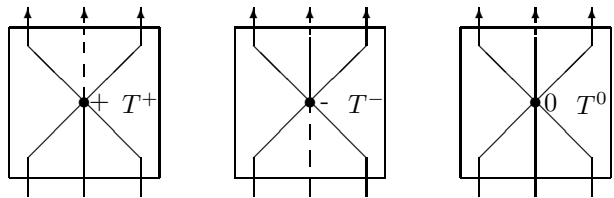
{\bf Figure 2.}Classes of triple intersections. \\

\unitlength=1.25mm
\special{em:linewidth 0.4pt}
\linethickness{0.4pt}
\begin{picture}(85.00,134.67)
\put(35.00,0.00){\line(0,1){10.22}}
\put(35.00,10.22){\line(1,1){10.00}}
\put(45.00,20.44){\line(0,1){9.78}}
\put(45.00,30.22){\line(0,1){9.78}}
\put(55.00,0.00){\line(0,1){20.00}}
\put(55.00,20.00){\line(-1,1){20.00}}
\put(35.00,40.00){\vector(0,1){20.00}}
\put(45.00,0.00){\line(0,1){10.22}}
\put(45.00,10.22){\line(-1,1){4.00}}
\put(38.00,16.89){\line(-1,1){3.00}}
\put(35.00,20.00){\line(1,1){20.00}}
\put(55.00,40.00){\line(-1,1){10.00}}
\put(45.00,50.22){\vector(0,1){9.78}}
\put(45.00,40.00){\line(1,1){4.00}}
\put(52.00,47.11){\line(1,1){3.00}}
\put(55.00,50.22){\vector(0,1){9.78}}
\put(5.00,0.00){\line(0,1){20.00}}
\put(5.00,20.00){\line(1,1){20.00}}
\put(25.00,40.00){\vector(0,1){20.00}}
\put(15.00,0.00){\vector(0,1){60.00}}
\put(25.00,0.00){\line(0,1){20.00}}
\put(25.00,20.00){\line(-1,1){20.00}}
\put(5.00,40.00){\vector(0,1){20.00}}
\put(65.00,0.00){\line(0,1){20.00}}
\put(65.00,20.00){\line(1,1){20.00}}
\put(85.00,40.00){\vector(0,1){20.00}}
\put(75.00,0.00){\line(0,1){10.22}}
\put(85.00,0.00){\line(0,1){10.22}}
\put(85.00,10.22){\line(-1,1){10.00}}
\put(75.00,20.44){\line(0,1){9.78}}
\put(75.00,30.22){\line(0,1){9.78}}
\put(75.00,10.22){\line(1,1){4.00}}
\put(81.00,16.89){\line(5,4){4.00}}
\put(85.00,20.00){\line(-1,1){20.00}}
\put(65.00,40.00){\line(1,1){10.00}}
\put(75.00,50.22){\vector(0,1){9.78}}
\put(75.00,40.00){\line(-1,1){4.00}}
\put(68.00,47.11){\line(-1,1){3.00}}
\put(65.00,50.22){\vector(0,1){9.78}}
\put(15.00,30.22){\makebox(0,0)[cc]{$\bullet$}}
\put(19.00,30.22){\makebox(0,0)[cc]{-}}
\put(45.00,30.22){\makebox(0,0)[cc]{$\bullet$}}
\put(50.00,30.22){\makebox(0,0)[cc]{+}}
\put(75.00,30.22){\makebox(0,0)[cc]{$\bullet$}}
\put(80.00,30.22){\makebox(0,0)[cc]{+}}
\put(30.00,30.22){\vector(1,0){5.00}}
\put(30.00,30.22){\vector(-1,0){5.00}}
\put(60.00,30.22){\vector(1,0){5.00}}
\put(60.00,30.22){\vector(-1,0){5.00}}
\put(11.00,73.78){\line(0,1){20.00}}
\put(11.00,93.78){\line(1,1){20.00}}
\put(31.00,113.78){\vector(0,1){20.00}}
\put(21.00,73.78){\vector(0,1){60.00}}
\put(31.00,73.78){\line(0,1){20.00}}
\put(31.00,93.78){\line(-1,1){20.00}}
\put(11.00,113.78){\vector(0,1){20.00}}
\put(21.00,104.00){\makebox(0,0)[cc]{$\bullet$}}
\put(25.00,104.00){\makebox(0,0)[cc]{-}}
\put(41.00,73.78){\vector(0,1){60.00}}
\put(51.00,73.78){\line(0,1){20.00}}
\put(51.00,93.78){\line(1,1){20.00}}
\put(71.00,113.78){\vector(0,1){20.00}}
\put(61.00,73.78){\vector(0,1){60.00}}
\put(71.00,73.78){\line(0,1){20.00}}
\put(71.00,93.78){\line(-1,1){20.00}}
\put(51.00,113.78){\vector(0,1){20.00}}
\put(61.00,104.00){\makebox(0,0)[cc]{$\bullet$}}
\put(81.00,73.78){\line(-1,1){8.00}}
\put(70.00,84.89){\line(-1,1){8.00}}
\put(60.00,94.67){\line(-1,1){3.00}}
\put(54.00,100.89){\line(-1,1){3.00}}
\put(51.00,104.00){\line(1,1){4.00}}
\put(57.00,109.78){\line(1,1){3.00}}
\put(60.00,112.89){\line(3,5){1.00}}
\put(63.00,116.89){\line(1,1){7.00}}
\put(73.00,126.67){\vector(1,1){8.00}}
\put(67.00,104.00){\makebox(0,0)[cc]{+,-,0}}
\put(6.00,104.00){\makebox(0,0)[cc]{and}}
\put(47.00,108.00){\vector(1,0){5.00}}
\put(47.00,108.00){\vector(-1,0){3.00}}
\end{picture}

\unitlength=1.25mm
\special{em:linewidth 0.4pt}
\linethickness{0.4pt}
\begin{picture}(72.00,70.22)
\put(50.00,0.00){\line(0,1){10.22}}
\put(50.00,10.22){\line(1,1){20.00}}
\put(70.00,30.22){\line(0,0){0.00}}
\put(70.00,30.22){\line(-2,1){20.00}}
\put(60.00,0.00){\line(0,1){10.22}}
\put(60.00,10.22){\line(-1,1){4.00}}
\put(54.00,16.00){\line(-1,1){4.00}}
\put(70.00,0.00){\line(0,1){10.22}}
\put(70.00,10.22){\line(-6,5){9.00}}
\put(50.00,20.00){\line(1,1){10.00}}
\put(60.00,30.22){\line(0,1){9.78}}
\put(59.00,20.89){\line(-1,1){3.00}}
\put(54.00,26.22){\line(-1,1){4.00}}
\put(50.00,30.22){\line(2,1){20.00}}
\put(60.00,50.22){\line(0,0){0.00}}
\put(70.00,40.00){\line(-1,1){20.00}}
\put(50.00,60.00){\vector(0,1){10.22}}
\put(60.00,40.00){\line(5,4){4.00}}
\put(67.00,46.22){\line(3,4){3.00}}
\put(70.00,50.22){\line(-1,1){10.00}}
\put(60.00,60.00){\vector(0,1){10.22}}
\put(50.00,40.00){\line(1,1){9.00}}
\put(61.00,51.11){\line(1,1){3.00}}
\put(67.00,56.89){\line(1,1){3.00}}
\put(70.00,60.00){\vector(0,1){10.22}}
\put(10.00,0.00){\line(0,1){30.22}}
\put(10.00,30.22){\line(2,1){20.00}}
\put(30.00,40.00){\vector(0,1){30.22}}
\put(30.00,0.00){\line(0,1){30.22}}
\put(30.00,30.22){\line(-2,1){20.00}}
\put(10.00,40.00){\vector(0,1){30.22}}
\put(20.00,0.00){\vector(0,1){70.22}}
\put(40.00,35.11){\vector(1,0){8.00}}
\put(40.00,35.11){\vector(-1,0){5.00}}
\put(20.00,35.11){\makebox(0,0)[cc]{$\bullet$}}
\put(29.00,35.11){\makebox(0,0)[cc]{0}}
\put(60.00,35.11){\makebox(0,0)[cc]{$\bullet$}}
\put(72.00,35.11){\makebox(0,0)[cc]{0}}
\end{picture}

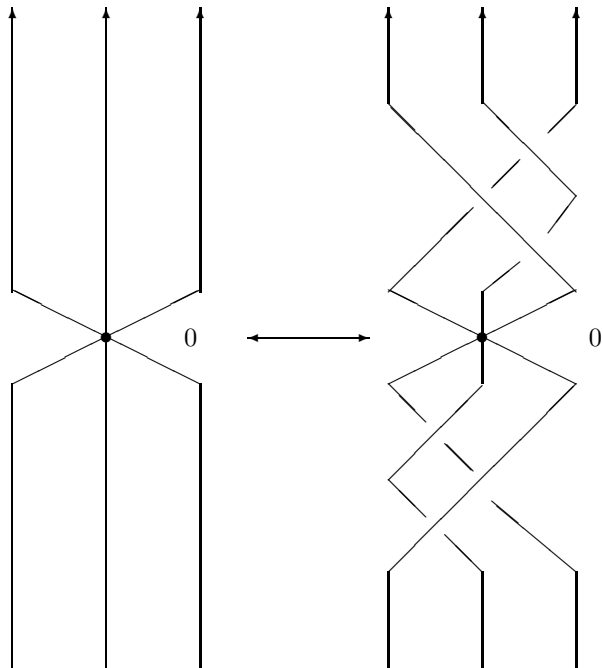
{\bf Figure 3.} Reidemeister moves for triple intersections.\\

\unitlength=0.50mm
\special{em:linewidth 0.4pt}
\linethickness{0.4pt}
\begin{picture}(115.00,55.11)
\put(50.00,4.89){\line(0,1){20.00}}
\put(50.00,24.89){\line(2,1){20.00}}
\put(70.00,35.11){\vector(0,1){20.00}}
\put(70.00,4.89){\line(0,1){20.00}}
\put(70.00,24.89){\line(-2,1){20.00}}
\put(50.00,35.11){\vector(0,1){20.00}}
\put(60.00,55.11){\vector(0,-1){50.22}}
\put(85.00,4.89){\line(0,1){15.11}}
\put(85.00,20.00){\line(3,2){30.00}}
\put(115.00,40.00){\vector(0,1){15.11}}
\put(115.00,4.89){\line(0,1){15.11}}
\put(115.00,20.00){\line(-3,2){30.00}}
\put(85.00,40.00){\vector(0,1){15.11}}
\put(105.00,4.89){\line(0,1){15.11}}
\put(105.00,20.00){\line(-1,2){10.00}}
\put(95.00,40.00){\vector(0,1){15.11}}
\put(105.00,55.11){\line(0,-1){15.11}}
\put(105.00,40.00){\line(-1,-2){10.00}}
\put(95.00,20.00){\vector(0,-1){15.11}}
\put(100.00,30.22){\makebox(0,0)[cc]{$\bullet$}}
\put(60.00,30.22){\makebox(0,0)[cc]{$\bullet$}}
\end{picture}
\\
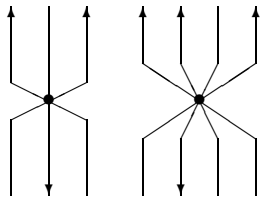
{\bf Figure 4.} Intersections in knots (links) that cannot be
expressed as
a closed braid.\\

\unitlength=0.80mm
\special{em:linewidth 0.4pt}
\linethickness{0.4pt}
\begin{picture}(110.00,55.11)
\put(10.00,15.11){\vector(0,1){40.00}}
\put(20.00,15.11){\line(0,1){9.78}}
\put(20.00,24.89){\line(1,1){20.00}}
\put(40.00,44.89){\vector(0,1){10.22}}
\put(40.00,15.11){\line(0,1){9.78}}
\put(40.00,24.89){\line(-1,1){20.00}}
\put(20.00,44.89){\vector(0,1){10.22}}
\put(50.00,15.11){\vector(0,1){40.00}}
\put(70.00,15.11){\vector(0,1){40.00}}
\put(80.00,15.11){\line(0,1){9.78}}
\put(80.00,24.89){\line(1,1){20.00}}
\put(100.00,44.89){\vector(0,1){10.22}}
\put(100.00,15.11){\line(0,1){9.78}}
\put(100.00,24.89){\line(-1,1){20.00}}
\put(80.00,44.89){\vector(0,1){10.22}}
\put(110.00,15.11){\vector(0,1){40.00}}
\put(90.00,15.11){\vector(0,1){40.00}}
\put(80.00,10.22){\makebox(0,0)[cc]{i-1}}
\put(90.00,10.22){\makebox(0,0)[cc]{i}}
\put(100.00,10.22){\makebox(0,0)[cc]{i+1}}
\put(75.00,35.11){\makebox(0,0)[cc]{....}}
\put(105.00,35.11){\makebox(0,0)[cc]{....}}
\put(90.00,35.11){\makebox(0,0)[lc]{$\bullet$ +,-,0}}
\put(60.00,35.11){\makebox(0,0)[cc]{and}}
\put(45.00,35.11){\makebox(0,0)[cc]{....}}
\put(30.00,35.11){\makebox(0,0)[cc]{$\bullet$}}
\put(15.00,35.11){\makebox(0,0)[cc]{....}}
\put(20.00,10.22){\makebox(0,0)[cc]{i}}
\put(40.00,10.22){\makebox(0,0)[cc]{i+1}}
\put(30.00,3.11){\makebox(0,0)[cc]{$a_i$}}
\put(90.00,3.11){\makebox(0,0)[cc]{$b^{\pm ,0}_i$}}
\end{picture}
\\
{\bf Figure 5.} Graphical representation of the generators of the
extended
braid algebra.\\

\unitlength=0.75mm
\special{em:linewidth 0.4pt}
\linethickness{0.4pt}
\begin{picture}(145.00,140.00)
\put(5.00,10.22){\vector(1,1){30.00}}
\put(35.00,10.22){\line(-1,2){9.00}}
\put(24.00,32.00){\vector(-1,2){4.00}}
\put(20.00,10.22){\line(-1,2){4.00}}
\put(14.00,22.22){\vector(-1,2){9.00}}
\put(45.00,10.22){\vector(1,2){15.00}}
\put(60.00,10.22){\vector(1,2){15.00}}
\put(75.00,10.22){\line(-1,1){9.00}}
\put(64.00,20.89){\line(-1,1){8.00}}
\put(54.00,31.11){\vector(-1,1){9.00}}
\put(85.00,10.22){\line(1,1){15.00}}
\put(100.00,10.22){\line(-1,1){7.00}}
\put(91.00,19.11){\line(-1,1){6.00}}
\put(85.00,24.89){\vector(1,1){15.00}}
\put(100.00,24.89){\line(-1,1){6.00}}
\put(91.00,34.22){\vector(-1,1){6.00}}
\put(130.00,10.22){\line(1,1){15.00}}
\put(145.00,10.22){\line(-1,1){7.00}}
\put(136.00,19.11){\line(-1,1){6.00}}
\put(130.00,24.89){\vector(1,1){15.00}}
\put(145.00,24.89){\line(-1,1){6.00}}
\put(136.00,34.22){\vector(-1,1){6.00}}
\put(110.00,10.22){\vector(0,1){29.78}}
\put(120.00,10.22){\vector(0,1){29.78}}
\put(10.00,60.00){\vector(1,3){10.00}}
\put(20.00,60.00){\line(-1,3){4.00}}
\put(14.00,78.22){\vector(-1,3){4.00}}
\put(30.00,60.00){\vector(0,1){30.22}}
\put(55.00,60.00){\vector(-1,3){10.00}}
\put(45.00,60.00){\line(1,3){4.00}}
\put(51.00,78.22){\vector(1,3){4.00}}
\put(65.00,60.89){\vector(0,1){29.33}}
\put(90.00,60.00){\vector(1,3){10.00}}
\put(100.00,60.00){\line(-1,3){4.00}}
\put(94.00,78.22){\vector(-1,3){4.00}}
\put(80.00,60.00){\vector(0,1){30.22}}
\put(135.00,60.00){\vector(-1,3){10.00}}
\put(125.00,60.00){\line(1,3){4.00}}
\put(131.00,78.22){\vector(1,3){4.00}}
\put(115.00,60.00){\vector(0,1){30.22}}
\put(45.00,110.22){\vector(0,1){29.78}}
\put(55.00,110.22){\vector(0,1){29.78}}
\put(65.00,110.22){\vector(0,1){29.78}}
\put(81.00,110.22){\vector(1,1){29.00}}
\put(95.00,110.22){\line(1,1){15.00}}
\put(110.00,125.33){\line(-1,1){6.00}}
\put(102.00,132.89){\vector(-1,1){7.00}}
\put(110.00,110.22){\line(-1,1){7.00}}
\put(101.00,119.11){\line(-1,1){5.00}}
\put(94.00,126.22){\vector(-1,1){13.00}}
\put(75.00,124.89){\makebox(0,0)[cc]{,}}
\put(95.00,100.00){\makebox(0,0)[cc]{$L_{2+}L_{1+}L_{2+}$}}
\put(55.00,100.00){\makebox(0,0)[cc]{$L_{10}L_{20}$}}
\put(20.00,50.22){\makebox(0,0)[cc]{$L_{1+}$}}
\put(38.00,75.11){\makebox(0,0)[cc]{,}}
\put(72.00,75.11){\makebox(0,0)[cc]{,}}
\put(106.00,75.11){\makebox(0,0)[cc]{,}}
\put(55.00,50.22){\makebox(0,0)[cc]{$L_{1-}$}}
\put(90.00,50.22){\makebox(0,0)[cc]{$L_{2+}$}}
\put(125.00,50.22){\makebox(0,0)[cc]{$L_{2-}$}}
\put(115.00,24.89){\makebox(0,0)[cc]{,}}
\put(80.00,24.89){\makebox(0,0)[cc]{,}}
\put(40.00,24.89){\makebox(0,0)[cc]{,}}
\put(20.00,0.00){\makebox(0,0)[cc]{$L_{2+}L_{1+}$}}
\put(60.00,0.00){\makebox(0,0)[cc]{$L_{1+}L_{2+}$}}
\put(100.00,0.00){\makebox(0,0)[cc]{$L_{1+}L_{1+}$}}
\put(130.00,0.00){\makebox(0,0)[cc]{$L_{2+}L_{2+}$}}
\end{picture}

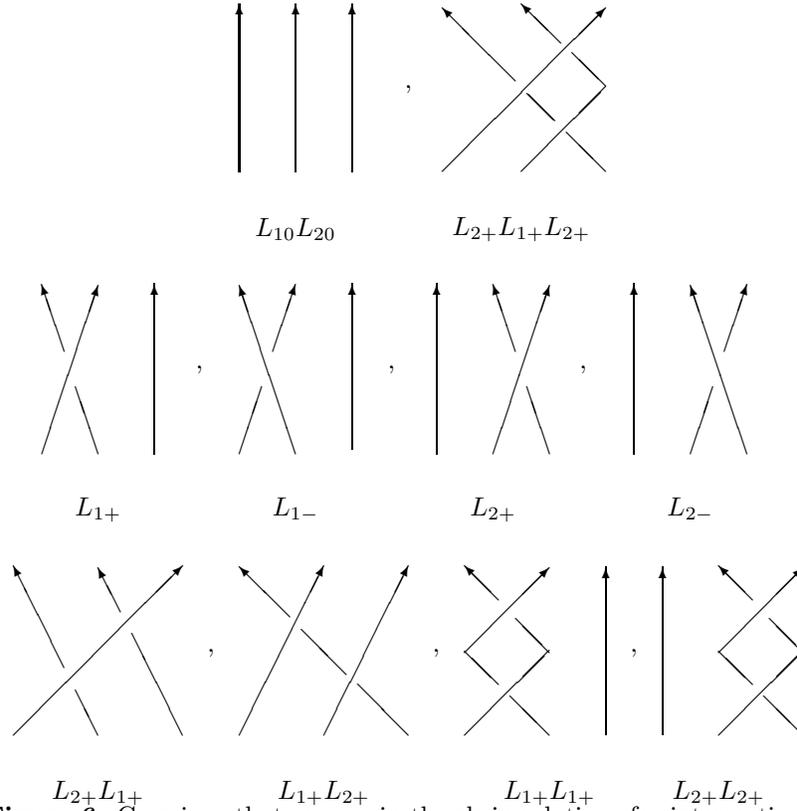
{\bf Figure 6.} Crossings that appear in the skein relations for
intersections.\\

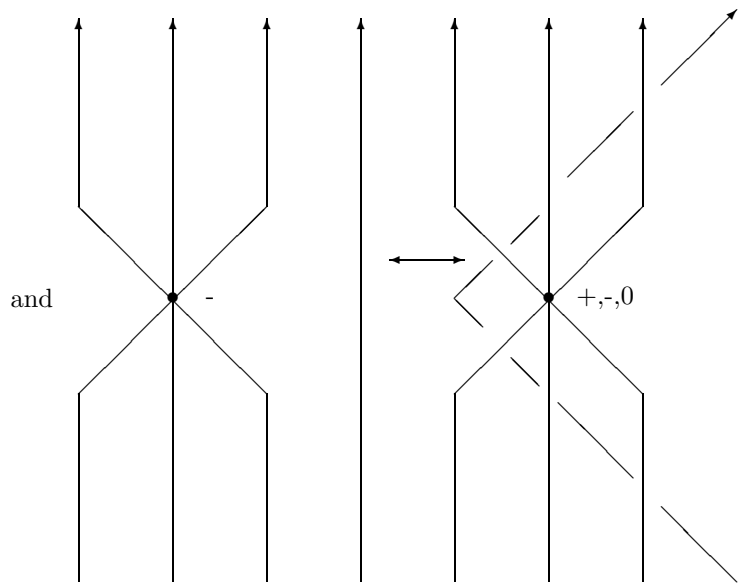
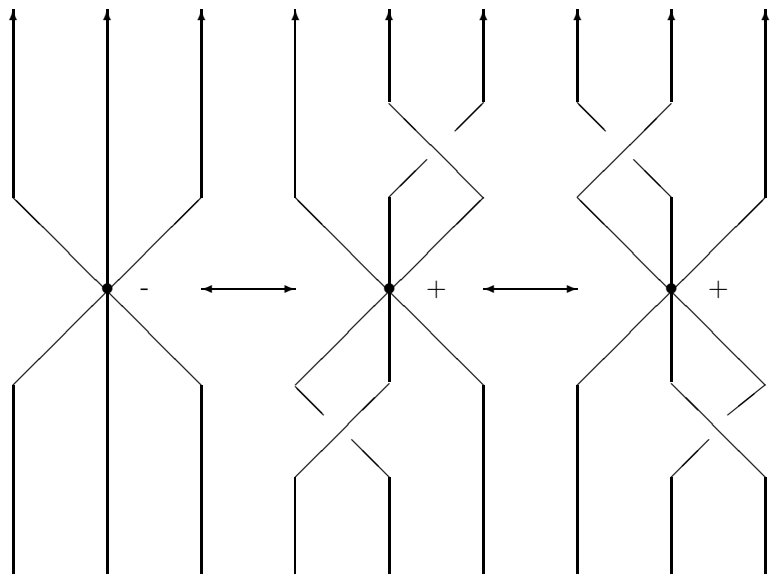
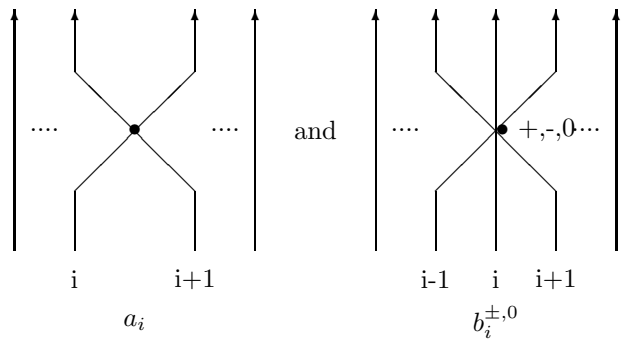
\end{document}